\documentclass[conference]{IEEEtran}
\IEEEoverridecommandlockouts
\usepackage[ruled,vlined,linesnumbered]{algorithm2e}
\usepackage{amsmath,amssymb,amsfonts}
\usepackage{amsthm}
\usepackage{array}
\usepackage{booktabs}
\usepackage{float}
\usepackage{graphicx}
\usepackage{mathrsfs}
\usepackage{multirow}
\usepackage{enumitem}
\usepackage{ragged2e}
\usepackage{setspace}
\usepackage{subfigure}
\usepackage{tabularx}
\usepackage{textcomp}
\usepackage{url}
\usepackage{verbatim}
\usepackage{xcolor}
\usepackage{xspace}
\usepackage[normalem]{ulem}
\usepackage{cite}
\newtheorem{theorem}{Theorem}

\newtheorem{definition}{Definition}

\AtBeginDocument{%
  \providecommand\BibTeX{{%
    \normalfont B\kern-0.5em{\scshape i\kern-0.25em b}\kern-0.8em\TeX}}}

\newcommand{\naive}{\texttt{Naive}\xspace}

\newcommand{\lapbit}{\texttt{Ind-Lap}\xspace}
\newcommand{\rrbit}{\texttt{Ind-RR}\xspace}

\newcommand{\dmax}{d_{max}\xspace}

\newcommand{\bcrifull}{Bounded Conditional Range Inversion\xspace}
\newcommand{\bcri}{BCRI\xspace}
\newcommand{\aesfull}{Adaptive Estimator Selection\xspace}
\newcommand{\aes}{AES\xspace}

\setlist[itemize]{label=\textbf{-},leftmargin=*}
\setlist[enumerate]{label=\textit{(\arabic*)},leftmargin=*}

\newcommand{\add}[1]{#1}

\newcommand{\limitedreps}[1]{\textcolor{black}{#1}}

\begin{document}
\title{\add{Privacy-Preserving Range Counting\\over Local Graph Statistics}}

\author{\IEEEauthorblockN{Gengda Zhao\textsuperscript{1}, Yizhang He\textsuperscript{1}\thanks{Yizhang He is the corresponding author.}, Dong Wen\textsuperscript{2}, Kai Wang\textsuperscript{1}, Ruijia Wu\add{\textsuperscript{3}}, Jianwei Wang\textsuperscript{2}}
\IEEEauthorblockA{\textsuperscript{1}Shanghai Jiao Tong University\quad\textsuperscript{2}University of New South Wales\quad\add{\textsuperscript{3}Zhejiang University}}
\IEEEauthorblockA{\add{\{gengda.zhao,yizhang.he,w.kai\}@sjtu.edu.cn}}
\IEEEauthorblockA{\{dong.wen,jianwei.wang1\}@unsw.edu.au\quad\add{rjwu@zju.edu.cn}}}

\maketitle

\begin{abstract}
Range counting over vertex statistics supports contact analysis, community profiling, and reputation screening. However, releasing exact range counts can disclose sensitive edges to an adversary with background knowledge. We study this problem under edge local differential privacy (edge LDP), where each vertex randomizes its report before sending it to an untrusted curator. We develop a layered framework that connects statistic estimation, direct range estimation, and adaptive estimator selection. \add{Statistic-estimation methods provide complementary scalar estimators with explicit loss analyses, while indicator-estimation methods privatize range indicators directly.} We next propose \bcrifull\ (\bcri), which derives unbiased range estimates from reusable bounded geometric reports through an explicit local inverse. Its boundary-conditioned estimator achieves no larger expected L2 loss than the unbounded inverse, and histogram preprocessing supports constant-time queries. \add{The same reports also provide a nonnegative, unbiased variance estimate for every answer.} At the center of the framework, \aesfull\ (\aes) privately predicts bias, variance, and covariance to combine \bcri with a conditional estimator from the same transcript. Experiments on diverse real-world graph tasks demonstrate the accuracy, efficiency, and workload scalability of the proposed framework.
\end{abstract}

\section{Introduction}
Range counting over vertex statistics is a fundamental problem in graph analysis. The statistic may be an ordinary degree, the number of neighbors carrying a public label, or the sum of signed ratings received or assigned by a vertex. These range counts capture connection volume, connection type, and reputation without releasing individual neighbor lists to the curator. They support social, biological, and financial network analysis \cite{zhang2016engagement,zhu2017geo,fang2017effective,fang2021cohesive,wang2021survey}. For example, authorities can count individuals in high-contact degree bands to allocate testing resources \cite{tan2025deeptrace,eubank2004modelling,trieu2020epione,chan2021pact}, while shifts in high-degree transaction ranges can flag suspicious activity \cite{pourhabibi2020fraud,wang2021survey,weber2018anti}. Public-label degrees distinguish connection types, and signed rating sums support reputation screening.

\begin{figure}[t]
\vspace*{-1em}
\centering\includegraphics[width=.98\columnwidth]{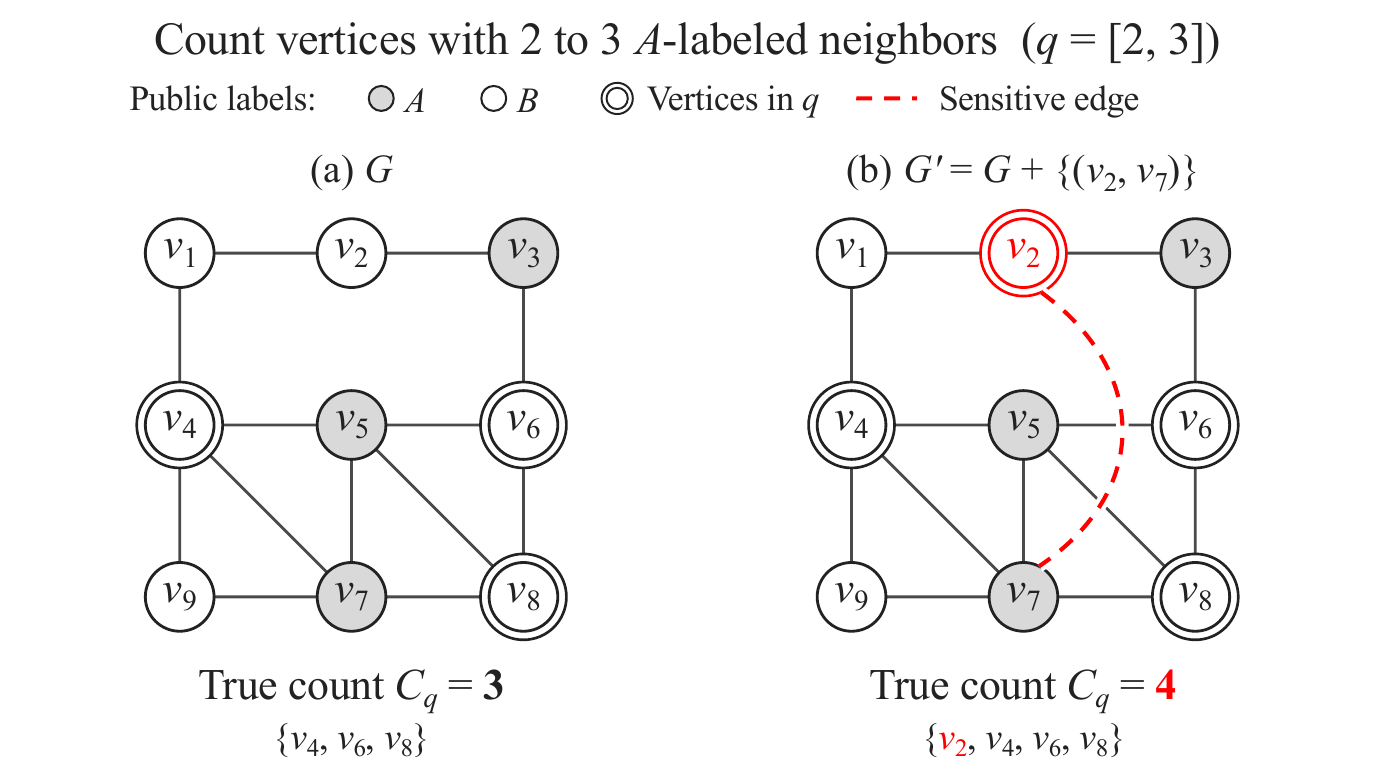}
\caption{Exact public-label range counts on two graphs differing by one sensitive edge.}
\vspace*{-1em}
\label{fig:example}
\end{figure}

Publishing these statistics without protection can reveal sensitive edges, so range queries need a privacy guarantee. Figure~\ref{fig:example} gives a concrete example. Shaded vertices carry public label $A$, and the query counts vertices whose numbers of $A$-labeled neighbors lie in the range $[2,3]$. The answer is three in graph (a) and four in graph (b). The only change is the dashed edge $(v_2,v_7)$, which moves $v_2$ into the query range by increasing its $A$-labeled neighbor count from one to two. An adversary who knows the other edges and public labels can use this change to infer the existence of that edge. Under local protection, each vertex randomizes its neighbor list or derived statistic before sending any report, so the curator never receives the exact local input.

\emph{Differential privacy} \cite{dwork2014algorithmic} is the standard framework for quantifying privacy loss. Classical central differential privacy (CDP) assumes a trusted curator who holds the full graph and releases noisy outputs \cite{dwork2014algorithmic,nissim2007smooth}, but a single breach of that curator can expose all users \cite{imola2021locally,jiang2021applications}. The \emph{local} model instead requires each participant to privatize their data before it leaves their control \cite{imola2021locally,imola2022communication}. For graphs, the natural sensitive units are often edges. \emph{Edge local differential privacy}, or edge LDP, requires each vertex to perturb its incident edges or derived vertex statistics before transmission to a curator \cite{qin2017generating,zhang2018two,ye2020lf}. For two neighbor lists differing by one edge, the privacy budget $\varepsilon$ bounds how much their report probabilities can differ. A smaller $\varepsilon$ provides stronger protection, while a larger $\varepsilon$ permits higher utility. 

This setting arises when participants hold their own contact or transaction records and a curator needs population-level range counts. Each vertex computes its statistic, randomizes a report, and sends it to the curator. Ordinary degree and public-label degree change by at most one under edge adjacency. A sum of fixed ratings bounded in absolute value by ten changes by at most ten. These public sensitivity bounds allow numerical local reports while protecting the underlying edges.

\noindent\textbf{Existing methods.}
Graph methods under edge LDP collect randomized adjacency or degree reports to estimate graph metrics~\cite{qin2017generating,zhang2018two,ye2020lf,imola2021locally,imola2022communication}. Applying randomized response (RR) to every potential edge creates a dense noisy graph. Debiasing its degrees corrects their expectations, but thresholding those estimates can still bias a range count. Direct Laplace perturbation has the same problem because an unbiased scalar estimate does not remain unbiased after a nonlinear range test. Privatizing the range indicator avoids thresholding bias, but each report is tied to one query and must share a fixed workload budget across predicates. General LDP range-query methods construct reusable hierarchical, wavelet, or numerical-distribution summaries~\cite{cormode2019range,du2021ahead,li2020squarewave}. For numerical estimation, Gu et al.~\cite{gu2019supporting} construct distance-based reporting probabilities and use probability-matrix inversion for unbiased frequency and range estimation. \add{Ghazi et al.~\cite{ghazi2022anonymized} construct anonymized histograms in shuffle and pan-private models by postprocessing discrete-Laplace-noised histograms. Hillebrand et al.~\cite{hillebrand2026privacy} show that discrete Laplace reports support unbiased estimation of subexponential functions and can be postprocessed into Laplace or Staircase outputs. These works establish general postprocessing principles, while we develop bounded range estimation, endpoint variance estimation, and adaptive combination under edge LDP.} These methods provide general reporting and inversion principles. Our setting calls for explicit estimators for clipped vertex-statistic channels, fast range evaluation, boundary-specific loss analysis, and private control of inverse-induced variance.

\noindent\textbf{Challenges.}
We face three main challenges.

\noindent\emph{$\bullet$ Challenge 1: Accurate statistic estimation with practical computation.}
Randomizing every potential edge creates a dense noisy graph and substantial computation, while direct scalar perturbation changes the probability that a vertex crosses a query endpoint. The challenge is to recover useful vertex statistics with explicit error and cost characterizations.

\noindent\emph{$\bullet$ Challenge 2: Direct and reusable range estimation.}
Indicator reports answer a range directly but remain tied to one predicate. A reusable numerical report must support arbitrary ranges, handle probability mass clipped at public-domain boundaries, and avoid domain-wide inversion or reconstruction.

\noindent\emph{$\bullet$ Challenge 3: Query-adaptive estimation with private information.}
The best bias--variance balance depends on the distribution around the query endpoints. The challenge is to predict and combine estimator losses using only privatized information, including the covariance induced by shared reports and the thresholding bias of a smoother estimator.

\noindent\textbf{Our approaches.}
We develop a layered framework spanning statistic estimation, direct range estimation, and adaptive estimator selection.

To address Challenge 1, Deg-RR debiases degrees from a shared randomized graph and characterizes the covariance induced by shared adjacency reports. Deg-Lap instead perturbs each vertex statistic directly, providing graph-size-independent scalar variance. Their exact L2 analyses explain both the computational trade-offs and the thresholding bias that subsequent range estimators must address.

To address Challenge 2, Ind-Lap and Ind-RR privatize the true range indicator, yielding unbiased query-specific estimates with sensitivity one. We then propose \bcrifull\ (\bcri) to make one bounded geometric transcript useful for arbitrary ranges. Its explicit local inverse avoids thresholding noisy statistics, its boundary-conditioned estimator removes uninformative tail randomness with a proved expected-L2-loss guarantee, and histogram preprocessing supports constant-time queries. \add{The same endpoint analysis yields a nonnegative, unbiased variance estimate from the received reports, allowing the curator to attach an error bar to every answer without additional privacy cost.}

At the center of the framework, \aesfull\ (\aes) addresses Challenge 3 by combining \bcri with Deg-Lap-RB, a conditional Laplace estimator derived from the same reports. \aes predicts bias, variance, and shared-report covariance to select a continuous combination weight for each query. Sensitivity-one statistics use full-budget cross-fitting, while bounded rating sums use a private pilot to compare the shared-report combination with Ind-RR.

\noindent\textbf{Contributions.}
\begin{itemize}
\item We formulate vertex-statistic range counting under edge LDP, covering ordinary degrees, public-label degrees, and signed rating sums through a common private-report interface.
\item We develop Deg-RR and Deg-Lap as complementary statistic estimators and derive their aggregate L2 losses, including shared-edge covariance and thresholding bias.
\item We develop Ind-Lap and Ind-RR as unbiased indicator estimators with sensitivity-one reports and explicit workload-budget costs.
\item We derive \bcri's explicit local inverse and boundary-conditioned estimator, prove unbiasedness and its expected-L2-loss guarantee, and support each query in $O(1)$ time after $O(n+B)$ preprocessing. \add{We further prove that endpoint report counts provide a nonnegative, unbiased variance estimate for every answer.}
\item We develop \aes, which combines shared-report estimators using privately predicted bias, variance, and covariance. We provide full-budget cross-fitting and private-pilot configurations and analyze their transcript privacy and complexity.
\item We conduct extensive experiments on diverse real-world graph tasks and established LDP baselines. The evaluation covers varying privacy budgets, query ranges, workload sizes, graph sizes, module ablations, running time, and communication, demonstrating the accuracy and efficiency of the proposed framework.
\end{itemize}

\section{Preliminaries}\label{sec:prelim}
\subsection{Graph Notation}
Let $G = (V, E)$ be an undirected graph with vertex set $V(G)$ and edge set $E(G)$, where $|V(G)| = n$ and $|E(G)| = m$.
The adjacency matrix of $G$ is denoted as $\mathcal{A}$, where $\mathcal{A}_{uv} = 1$ if $(u,v) \in E(G)$ and $\mathcal{A}_{uv} = 0$ otherwise. The $u$-th row of $\mathcal{A}$ represents the \emph{neighbor list} of $u$, denoted by $\mathcal{A}_u$. We use $N(u,G)$ to represent the neighbors of a vertex $u$ in $G$. The vertex degree is $d(u,G)=|N(u,G)|$.
Beyond this binary undirected special case, $\mathcal A_u$ denotes a weighted directed neighbor list when ratings are present, with the task selecting incoming or outgoing edges. Each vertex computes a raw integer statistic $\widetilde s_u=\phi(\mathcal A_u)$ from its neighbor list and public task attributes. A public shift and clipping map $\widetilde s_u$ to the encoded statistic $s_u\in\{0,\ldots,B\}$. Examples of $\widetilde s_u$ are degree, public-label degree, and incoming or outgoing signed rating sum. The sensitivity $\Delta$ is a public bound on $|\widetilde s_u-\widetilde s'_u|$ when one incident edge is added or removed. It is one for the two count statistics and ten for ratings with absolute value at most ten. Public shifting and clipping are nonexpansive, so $|s_u-s'_u|\leq\Delta$ also holds for the encoded statistics.

\begin{definition}[Range count]\label{def:rangequery}
For an integer range $q=[x,y]\subseteq[0,B]$, let $f_q(s)=\mathbf1\{x\leq s\leq y\}$. The answer is
$C_q=\sum_{u\in V}f_q(s_u)$.
A workload $\mathcal Q$ contains $Q$ public range queries. \add{Assume $B\geq1$, $\Delta>0$, $\varepsilon>0$, and nonempty $\mathcal Q$.}
\end{definition}
When a range includes a public-domain boundary code, it includes the tail probability mass accumulated at that boundary cell by clipping. Query interpretation therefore follows the public encoding. \add{The public schema specifies the vertex roster, labels, sensitivity bound, domain, and workload.}

\subsection{Differential Privacy for Graph Analysis}
Classic differential privacy assumes a trusted curator that accesses the original data~\cite{dwork2014algorithmic}. Edge LDP instead protects a vertex's adjacency list before its release~\cite{qin2017generating,imola2021locally}.
\begin{definition}[$\varepsilon$-edge LDP]\label{def:ldp}
A local randomizer $R_u$ is $\varepsilon$-edge locally differentially private if, for any adjacency lists $a,a'$ differing by one edge addition or removal and every measurable output set $S$,
\begin{equation}\label{eq:ldp}
\Pr[R_u(a)\in S]\leq e^\varepsilon\Pr[R_u(a')\in S].
\end{equation}
For interactive protocols, the guarantee applies to the vertex's complete transcript, including responses conditional on previous messages.
\end{definition}
This definition supports both noisy scalar reports and randomized adjacency. For signed data, edge addition and removal carry a fixed rating bounded by ten.

The guarantee in Definition~\ref{def:ldp} applies to each adjacency list. If both endpoints of an undirected edge report independently with budget $\varepsilon$, their joint graph-edge budget is $2\varepsilon$. Allocating $\varepsilon_G/2$ to each vertex therefore achieves a graph-wide budget $\varepsilon_G$.

\noindent\textbf{Problem statement.}
Estimate $C_q$ for each $q\in\mathcal Q$ from locally randomized reports, minimizing the expected L2 loss $\mathcal L(\widehat C_q)=\mathbb E[(\widehat C_q-C_q)^2]$,
subject to a stated per-vertex transcript budget. We report two settings: budget $\varepsilon$ per query, and budget $\varepsilon$ for an entire workload. A single reusable scalar report costs $\varepsilon$ for any number of postprocessed answers. $Q$ separately randomized predicates require composition. \add{Expectations are over mechanism randomness with fixed inputs.}

\subsection{A Naive Approach}
Binary randomized response (RR) flips a bit with probability $\pi=(1+e^\varepsilon)^{-1}$~\cite{warner1965randomized}. The debiased report $(Z-\pi)/(1-2\pi)$ is unbiased. A scalar statistic of sensitivity $\Delta$ can instead be released with Laplace noise of scale $\Delta/\varepsilon$. Postprocessing a private transcript costs no additional budget. Sequentially collecting reports with conditional budgets $\varepsilon_0$ and $\varepsilon_1$ costs at most their sum.

Applying RR to adjacency bits and directly counting vertices whose noisy degrees lie in the range gives a simple \naive approach. False-positive edges make the randomized graph dense, substantially shifting its degree distribution. Full simulation requires $\Theta(n^2)$ RR operations and one-time noisy-graph communication, after which the noisy degrees can be reused. We use this observation to motivate the reference estimators below.

\section{Range-Query Answering Based on Statistic Estimation}\label{sec:degree}
The naive construction in Section~\ref{sec:prelim} thresholds degrees in a randomized graph. With flip probability $\pi=(1+e^\varepsilon)^{-1}$, the expected noisy degree is $\pi(n-1)+(1-2\pi)d_u$, so false-positive edges shift a sparse graph toward a much denser degree distribution. A natural improvement first estimates each statistic and then applies the range test. Deg-RR and Deg-Lap instantiate this approach, and their exact loss analyses expose the thresholding term addressed by \bcri and \aes.

\subsection{Deg-RR: Randomized Response for Degree Range Query}
\noindent\textbf{Unbiased statistic estimation.}
Deg-RR assigns each unordered pair $\{u,v\}$ to one public vertex, which randomizes its adjacency bit once. The curator uses the shared report $Z_{uv}=Z_{vu}$ at both vertices, avoiding two independently randomized versions of the same edge~\cite{hillebrand2024cycle,eden2023triangle}. Since $\mathbb E Z_{uv}=\pi+(1-2\pi)\mathcal A_{uv}$, the report $\phi_{uv}=(Z_{uv}-\pi)/(1-2\pi)$ is unbiased for $\mathcal A_{uv}$. With $D'_u=\sum_{v\ne u}Z_{uv}$, the degree estimate is $\widehat d_u=[D'_u-(n-1)\pi]/(1-2\pi)$. Reports incident to a fixed vertex are independent, giving $\mathbb E\widehat d_u=d_u$ and $\operatorname{Var}(\widehat d_u)=(n-1)\pi(1-\pi)/(1-2\pi)^2$. This variance grows with the number of possible neighbors, including nonneighbors.

\noindent\textbf{Range estimation and exact L2 loss.}
The curator returns $\widehat C_{\rm DRR}=\sum_u I_u$, where $I_u=\mathbf1\{x\leq\widehat d_u\leq y\}$. Affine correction makes each degree estimate unbiased, but the nonlinear threshold can still move it across a query endpoint. For a fixed graph, $D'_u$ is the sum of independent $\operatorname{Bin}(d_u,1-\pi)$ and $\operatorname{Bin}(n-1-d_u,\pi)$ variables. Let $p_u=\Pr[I_u=1]$ be its exact hit probability and $b=\sum_u p_u-C_q$. Then
\begin{equation}\label{eq:rrcov}
\mathcal L(\widehat C_{\rm DRR})=
 b^2+\sum_u p_u(1-p_u)+2\sum_{u<v}\operatorname{Cov}(I_u,I_v).
\end{equation}
The covariance terms arise because $I_u$ and $I_v$ share the report $Z_{uv}$. Conditional on that report, the remaining randomized edges are disjoint, so the two indicators are independent. Their covariance is therefore determined by the shared report and can have either sign. Equation~\eqref{eq:rrcov} retains every pairwise term in the exact aggregate L2 loss.

\begin{theorem}
Deg-RR satisfies $\varepsilon$-edge LDP.
\end{theorem}
\begin{proof}
Changing one owned adjacency bit changes one binary RR input, whose likelihood ratio is at most $e^\varepsilon$. Degree correction and thresholding are postprocessing.
\end{proof}

\noindent\textbf{Computational time complexity.}
Full simulation randomizes $\Theta(n^2)$ pairs. Once degrees are accumulated, sorting answers $Q$ ranges in $O(n\log n+Q\log n)$ time without further reports.

\noindent\textbf{Space complexity and per-range communication costs.}
Degree accumulation uses $O(n)$ curator space. A dense report transmits $\Theta(n^2)$ bits, while a positive-edge list uses $O(m'\log n)$ bits for $m'$ reported edges. The accumulated degrees are reused across queries.

\subsection{Deg-Lap: Laplace Mechanism for Statistic Range Query}
\noindent\textbf{Unbiased statistic estimation.}
Deg-Lap avoids the noisy graph. Each vertex sends $Y_u=s_u+L_u$, where $L_u\sim\operatorname{Lap}(\Delta/\varepsilon)$ independently. Before clipping, $\mathbb E Y_u=s_u$ and $\operatorname{Var}(Y_u)=2\Delta^2/\varepsilon^2$. For sensitivity-one statistics, this variance is independent of graph size, unlike Deg-RR.

\noindent\textbf{Range estimation and exact L2 loss.}
An integer range includes the entire cells of its endpoint values. Deg-Lap therefore tests $a_q\leq Y_u\leq b_q$, with $a_q=x-1/2$ and $b_q=y+1/2$, and returns $\widehat C_L=\sum_u\mathbf1\{a_q\leq Y_u\leq b_q\}$. The lower limit is $-\infty$ when $x=0$, and the upper limit is $+\infty$ when $y=B$, so public-domain boundary codes include tail probability mass accumulated by clipping. Directly using $[x,y]$ as a continuous range would give an interior singleton query zero hit probability.

Let $\eta=\varepsilon/\Delta$ and $F_\eta$ be the Laplace CDF, namely $F_\eta(z)=e^{\eta z}/2$ for $z<0$ and $1-e^{-\eta z}/2$ otherwise. The hit probability at true value $s$ is $p_L(s)=F_\eta(b_q-s)-F_\eta(a_q-s)$. Independence across vertices gives $\mathcal L(\widehat C_L)=\sum_u p_L(s_u)(1-p_L(s_u))+[\sum_u(p_L(s_u)-f_q(s_u))]^2$. The first term can be small in a sparse tail because false positives are rare. Near a populated endpoint, however, signed membership errors accumulate in the squared-bias term. For an interior singleton $[x,x]$, a vertex with $s_u=x$ is counted with probability $1-e^{-\eta/2}$, and many such vertices reinforce the same negative bias. Unbiased scalar perturbation therefore does not imply an unbiased range count.

\begin{theorem}
Deg-Lap satisfies $\varepsilon$-edge LDP.
\end{theorem}
\begin{proof}
For $|s-s'|\leq\Delta$, the Laplace density ratio is at most $\exp(\varepsilon|s-s'|/\Delta)\leq e^\varepsilon$. Subsequent thresholding and public clipping are postprocessing.
\end{proof}

\noindent\textbf{Computational time complexity.}
The curator can sort the $O(n)$ reports once and answer range queries by rank lookups, taking $O(n\log n+Q\log n)$ time for a workload.

\noindent\textbf{Space complexity and per-range communication costs.}
The curator stores $O(n)$ scalar reports. Each vertex sends one scalar for the entire workload. The thresholding term leads naturally to indicator estimation and the direct bounded-channel inversion in Section~\ref{sec:bcri}.

\section{Range-Query Answering via Indicator Estimation}\label{sec:indicator}
The previous methods estimate a statistic before applying a discontinuous range test. Indicator methods reverse these operations. Each vertex computes the true range indicator $b_u=f_q(s_u)$ and privatizes that bit. One incident edge can change $b_u$ by at most one, even when the numerical statistic has sensitivity $\Delta>1$. \add{This removes thresholding bias and scalar sensitivity from local randomization.}

\subsection{\lapbit: Laplace Perturbation of Indicators}
\noindent\textbf{Indicator estimation and exact L2 loss.}
Vertex $u$ sends $X_u=b_u+L_u$ with independent $L_u\sim\operatorname{Lap}(1/\varepsilon)$, and the curator returns $\widehat C_{\rm IL}=\sum_uX_u$ without another threshold. Zero-mean noise gives $\mathbb E\widehat C_{\rm IL}=C_q$ and $\mathcal L(\widehat C_{\rm IL})=2n/\varepsilon^2$. Every vertex contributes the same variance regardless of its distance from a query endpoint. The estimate is therefore stable across query locations, and its expected L2 loss does not decrease when statistic mass near the endpoints is small.

\begin{algorithm}[t]
\caption{Ind-Lap}\label{alg:ind-lap}
\KwIn{Range query $q$ and privacy budget $\varepsilon$}
\KwOut{Estimated range count $\widehat C_{\rm IL}$}
\tcp{Vertex side}
\ForEach{vertex $u$}{
Compute the true indicator $b_u=f_q(s_u)$\;
Sample $L_u\sim\operatorname{Lap}(1/\varepsilon)$\;
Send $X_u=b_u+L_u$ to the curator\;
}
\tcp{Curator side}
Return $\widehat C_{\rm IL}=\sum_uX_u$\;
\end{algorithm}

\noindent\textbf{The Ind-Lap algorithm.}
Algorithm~\ref{alg:ind-lap} privatizes the true predicate before any curator computation. Each vertex releases only the noisy indicator $X_u$, whose zero-mean noise allows direct unbiased aggregation. The resulting scalar report encodes the public query $q$.

\begin{theorem}
Ind-Lap satisfies $\varepsilon$-edge LDP.
\end{theorem}
\begin{proof}
\add{The sensitivity-one indicator gives $\varepsilon$-edge LDP, and aggregation is postprocessing.}
\end{proof}

\noindent\textbf{Computational time complexity.}
Reporting and aggregation take $O(n)$ time for each query.

\noindent\textbf{Space complexity and per-range communication costs.}
Each vertex sends one scalar per query, and summation uses constant curator working space.

\subsection{\rrbit: Randomized Response on Indicators}
\noindent\textbf{Indicator estimation and exact L2 loss.}
Let $\pi=(1+e^\varepsilon)^{-1}$. A vertex reports the true bit with probability $1-\pi$ and flips it otherwise. The curator returns $\widehat C_{\rm IR}=\sum_u(Z_u-\pi)/(1-2\pi)$. Since $\mathbb E Z_u=\pi+(1-2\pi)b_u$, each transformed report and the resulting count are unbiased. Independent reports have $\operatorname{Var}(Z_u)=\pi(1-\pi)$, giving
\begin{equation}\label{eq:rr-risk}
R_{\rm RR}(n,\varepsilon)=\mathcal L(\widehat C_{\rm IR})
=\frac{ne^\varepsilon}{(e^\varepsilon-1)^2}
=\frac{n}{4\sinh^2(\varepsilon/2)}.
\end{equation}
For fixed $n$ and $\varepsilon$, this expected L2 loss is independent of the statistic distribution and range count. Because $\sinh(t)>t$ for $t>0$, it is below half the Ind-Lap expected L2 loss.

\begin{algorithm}[t]
\caption{Ind-RR}\label{alg:ind-rr}
\KwIn{Range query $q$ and privacy budget $\varepsilon$}
\KwOut{Estimated range count $\widehat C_{\rm IR}$}
\tcp{Vertex side}
\ForEach{vertex $u$}{
Compute the true indicator $b_u=f_q(s_u)$\;
Report $Z_u=b_u$ with probability $1-\pi$, otherwise report $1-b_u$\;
}
\tcp{Curator side}
Return $\widehat C_{\rm IR}=\sum_u(Z_u-\pi)/(1-2\pi)$\;
\end{algorithm}

\noindent\textbf{The Ind-RR algorithm.}
Algorithm~\ref{alg:ind-rr} uses a single randomized bit per vertex and query. The curator applies the affine correction $(Z_u-\pi)/(1-2\pi)$ before aggregation, yielding an unbiased range count directly from the binary reports.

\begin{theorem}
Ind-RR satisfies $\varepsilon$-edge LDP.
\end{theorem}
\begin{proof}
\add{The indicator likelihood ratio is $e^\varepsilon$. Debiasing and aggregation are postprocessing.}
\end{proof}

\noindent\textbf{Computational time complexity.}
Reporting, debiasing, and aggregation take $O(n)$ time for each query.

\noindent\textbf{Space complexity and per-range communication costs.}
Each vertex sends one bit per query, and the curator can aggregate reports in streaming fashion.

\noindent\textbf{Complementarity and workload cost.}
Numerical thresholding can accumulate signed errors near populated endpoints, whereas indicator estimation retains the exact expected L2 loss in~\eqref{eq:rr-risk}. For a query range with little statistic mass, Deg-Lap can instead benefit from rare false positives and low hit-probability variance. This complementarity motivates adaptive estimation, but an indicator report is tied to its public predicate. Answering $Q$ distinct queries with fresh reports under total budget $\varepsilon$ allocates $\varepsilon/Q$ to each query, giving expected L2 losses $R_{\rm RR}(n,\varepsilon/Q)$ and $2nQ^2/\varepsilon^2$ for Ind-RR and Ind-Lap. \bcri instead provides unbiased range estimates from a reusable numerical transcript.

\section{\bcrifull: Inverting the Bounded Geometric Channel}\label{sec:bcri}
The preceding sections leave two complementary problems. Deg-Lap provides a reusable numerical report but incurs thresholding bias, while Deg-RR additionally requires a costly randomized graph with shared-edge dependence. Ind-Lap and Ind-RR avoid thresholding bias, but their predicate-specific reports require fresh privacy budget for distinct queries and give every vertex the same noise scale regardless of its distance from an endpoint. They therefore cannot use the numerical locality of a graph statistic.

Distance-based reporting and probability-matrix inversion provide a general route to unbiased local estimation~\cite{gu2019supporting}. \bcri develops this route for bounded vertex statistics through three specific designs: an explicit local inverse for each range indicator, histogram and prefix structures for constant-time query evaluation, and a bounded conditional estimator for probability mass accumulated at public-domain boundaries. The curator directly estimates the range indicator from one reusable geometric report per vertex.

\subsection{The \bcri Framework}
Let $\alpha=\exp(-\varepsilon/\Delta)$. Each vertex samples two-sided geometric noise $Z_u$ with
\begin{equation}\label{eq:geom}
\Pr[Z_u=z]=\frac{1-\alpha}{1+\alpha}\alpha^{|z|},\quad z\in\mathbb Z,
\end{equation}
and sends $T_u=[s_u+Z_u]_0^B$, where $[v]_0^B=\min(B,\max(0,v))$. The geometric mechanism is a standard integer-noise primitive~\cite{ghosh2012geometric}. Algorithm~\ref{alg:bcri} gives the complete reporting and query procedure, using the inverse weights derived below.

\begin{algorithm}[t]
\caption{\bcri}\label{alg:bcri}
\KwIn{Public $B,\Delta,\varepsilon$ and range workload $\mathcal Q$}
\KwOut{Estimated range counts $\widehat C_q$, $q\in\mathcal Q$}
\tcp{Vertex side}
\ForEach{vertex $u$}{
Compute the encoded local statistic $s_u\in\{0,\ldots,B\}$\;
Sample $Z_u$ according to the distribution in~\eqref{eq:geom}\;
Send $T_u=[s_u+Z_u]_0^B$ to the curator\;
}
\tcp{Curator side}
Build report histogram $H(t)=|\{u:T_u=t\}|$\;
Build prefix counts $A(j)=\sum_{t<j}H(t)$\;
\ForEach{$q=[x,y]\in\mathcal Q$}{
Obtain $g_q$ from~\eqref{eq:geo-weights}\;
Compute $\widehat C_q=\sum_t H(t)g_q(t)$ using a prefix range sum and public-domain boundary corrections\;
Return $\widehat C_q$\;
}
\end{algorithm}

\noindent\textbf{Reusable query evaluation.}
Each vertex contributes one clipped integer, and the curator compacts all reports into the histogram $H$. The inverse in~\eqref{eq:geo-weights} equals one inside a nontrivial query range and zero away from it, apart from a constant number of cells near $x,y,0,B$. Each query therefore requires one prefix range sum and these local corrections. This design supports arbitrary later queries from the same transcript without materializing a domain-wide weight vector.

\noindent\textbf{Computational time complexity.}
Building $H$ and $A$ costs $O(n+B)$ time, after which each range query takes constant time and the full workload takes $O(Q)$ time.

\noindent\textbf{Space complexity and per-range communication costs.}
The histogram and prefix structures use $O(B)$ space. Each vertex sends one integer, and all queries reuse the same transcript without additional per-range reports.

\subsection{The Bounded Reporting Channel}
The report distribution is $P_{st}=\Pr[T_u=t\mid s_u=s]$, where
\begin{equation}\label{eq:bounded-channel}
P_{st}=\begin{cases}
\alpha^s/(1+\alpha),&t=0,\\
\dfrac{1-\alpha}{1+\alpha}\alpha^{|s-t|},&0<t<B,\\
\alpha^{B-s}/(1+\alpha),&t=B.
\end{cases}
\end{equation}
The public-domain boundary cells collect out-of-domain tail probability mass. Treating them as ordinary interior cells would invert a different channel.

\begin{theorem}\label{thm:geo-privacy}
The report $T_u$ satisfies $\varepsilon$-edge LDP for any statistic of sensitivity at most $\Delta$. All range-query answers computed from this report have the same transcript guarantee.
\end{theorem}
\begin{proof}
For neighboring statistics $s,s'$, the likelihood ratio of the unbounded report is at most $\alpha^{-|s-s'|}\leq e^\varepsilon$ by the triangle inequality. Clipping and subsequent estimation are postprocessing. The same bound follows directly from the boundary probabilities in~\eqref{eq:bounded-channel}.
\end{proof}

\subsection{Inverse Weights and Unbiased Counts}
For a query indicator $f=f_q$, we derive weights $g_q$ satisfying $Pg_q=f$. Writing $f_t=f_q(t)$, the inverse is local:
\begin{equation}\label{eq:geo-weights}
g_q(t)=\begin{cases}
\dfrac{f_0-\alpha f_1}{1-\alpha},&t=0,\\[4pt]
\dfrac{(1+\alpha^2)f_t-\alpha f_{t-1}-\alpha f_{t+1}}{(1-\alpha)^2},&0<t<B,\\[4pt]
\dfrac{f_B-\alpha f_{B-1}}{1-\alpha},&t=B.
\end{cases}
\end{equation}
\bcri returns $\widehat C_G=\sum_u g_q(T_u)$. Negative weights near a query endpoint cancel expected false positives from neighboring values, while weights above one compensate for false negatives. This cancellation operates in expectation and requires no estimate of the population distribution.

\begin{theorem}\label{thm:geo-unbiased}
For every fixed collection of encoded statistics, $\mathbb E\widehat C_G=C_q$. With independent vertex randomization,
\begin{equation}\label{eq:geo-risk}
\mathcal L(\widehat C_G)=\sum_u\left[\sum_{t=0}^B P_{s_ut}g_q(t)^2-f_q(s_u)\right].
\end{equation}
\end{theorem}
\begin{proof}
Multiplying~\eqref{eq:bounded-channel} by~\eqref{eq:geo-weights} gives $Pg_q=f_q$. Consecutive powers of $\alpha$ cancel in the interior, and the boundary rows account for out-of-domain tail probabilities accumulated at $0$ and $B$. Thus $\mathbb E[g_q(T_u)\mid s_u]=f_q(s_u)$, including singleton and boundary-touching queries. Summing over vertices and using independence and $f_q^2=f_q$ yields~\eqref{eq:geo-risk}.
\end{proof}

\subsection{\add{Endpoint Form and Variance Estimation}}
\add{Let $H(t)=|\{u:T_u=t\}|$ be the number of reports equal to $t$, $N_q=\sum_{t=x}^{y}H(t)$ be the report count inside the query range, and $c_\alpha=\alpha/(1-\alpha)^2$. The interior row of~\eqref{eq:geo-weights} becomes $g_q(t)=f_t-c_\alpha(f_{t-1}+f_{t+1}-2f_t)$. Hence corrections are nonzero only where $f_q$ changes. For $2\leq x\leq y\leq B-2$,}
\begin{equation}\label{eq:endpoint-form}
\add{\widehat C_G=N_q+c_\alpha\left[(H(x)-H(x-1))+(H(y)-H(y+1))\right].}
\end{equation}
\add{The range count is corrected by the net report flow across the two query endpoints, and all remaining report counts have weight zero or one.}

\begin{theorem}\label{thm:geo-variance}
\add{For every range $q$ and every fixed collection of encoded statistics,}
\begin{equation}\label{eq:endpoint-variance}
\add{\mathcal L(\widehat C_G)=\sum_{t=0}^{B}\mathbb E[H(t)]\,g_q(t)\bigl(g_q(t)-1\bigr).}
\end{equation}
\add{Every coefficient $g_q(t)(g_q(t)-1)$ is nonnegative and vanishes unless $t\in\{x-1,x,y,y+1\}$. Consequently, $\widehat V_q=\sum_t H(t)\,g_q(t)(g_q(t)-1)$ is a nonnegative unbiased estimate of $\mathcal L(\widehat C_G)$. For $2\leq x<y\leq B-2$, the coefficient is $c_\alpha(1+c_\alpha)$ on each of the four cells.}
\end{theorem}
\begin{proof}
\add{Theorem~\ref{thm:geo-unbiased} gives $\mathbb E[g_q(T_u)\mid s_u]=f_q(s_u)$ and $f_q^2=f_q$, so $\operatorname{Var}(g_q(T_u))=\mathbb E[g_q(T_u)(g_q(T_u)-1)]$. Independence and grouping by report value give~\eqref{eq:endpoint-variance}. If $f_{t-1}=f_t=f_{t+1}$, then $g_q(t)=f_t\in\{0,1\}$ and the coefficient vanishes. Interior and boundary weights satisfy $g_q(t)\leq0$ or $g_q(t)\geq1$, proving nonnegativity. For $2\leq x<y\leq B-2$, the weights are $-c_\alpha$ at $x-1$ and $y+1$ and $1+c_\alpha$ at $x$ and $y$. Replacing $\mathbb E[H(t)]$ by $H(t)$ preserves the expectation.}
\end{proof}
\add{Thus $\widehat V_q$ reads at most four report counts from the existing transcript and provides an error bar $\widehat C_G\pm z\widehat V_q^{1/2}$ without additional privacy cost. Section~\ref{sec:paired} evaluates this estimate.}

\subsection{Boundary Conditioning}
For comparison, define the unbounded inverse $g_q^\infty(z)$ by the interior expression in~\eqref{eq:geo-weights} for all $z\in\mathbb Z$, extending $f_q$ by zero outside $[0,B]$. It is unbiased for the unbounded report $W_u=s_u+Z_u$.
\begin{theorem}\label{thm:geo-rb}
\add{For each $s\in[0,B]$, $g_q(T)=\mathbb E[g_q^\infty(W)\mid T]$, and $\mathcal L(\widehat C_G)\leq\mathcal L\left(\sum_u g_q^\infty(W_u)\right)$.}
\end{theorem}
\begin{proof}
For $0<T<B$, $W=T$. Conditional on $T=0$, $W=-k$ has probability $(1-\alpha)\alpha^k$ for $k\geq0$, independently of $s\geq0$. Substituting $g_q^\infty$ gives the first row of~\eqref{eq:geo-weights}, and the upper tail is symmetric. The law of total variance proves the claim for each vertex and hence for their independent sum.
\end{proof}

\add{Theorem~\ref{thm:geo-variance} makes numerical locality precise: only reports adjacent to query endpoints contribute variance, unlike the population-wide indicator loss in~\eqref{eq:rr-risk}. When $\varepsilon/\Delta$ is small, $c_\alpha(1+c_\alpha)$ grows and motivates the bias--variance adaptation developed by \aes in Section~\ref{sec:aes}.}

\section{\aesfull}\label{sec:aes}
\bcri provides unbiased range estimates, with inverse amplification governed by $\varepsilon/\Delta$ and aggregate variance governed by the statistic mass near query endpoints. A smoother estimator offers complementary behavior through lower variance and controlled thresholding bias. \aes exploits this complementarity to reduce query-specific error. Its key idea is to combine estimators derived from the same geometric transcript using privately predicted L2 loss, including the covariance created by sharing those reports.

\subsection{The \aes Framework}
\aes uses full-budget cross-fitting for sensitivity-one statistics. Each vertex sends one bounded geometric report, and a public hash partitions the vertices into two folds. For fold $k$, let $G_{k,q}$ and $L_{k,q}$ be its \bcri and conditional Laplace estimates for query $q$, with weight $\lambda_{k,q}$ assigned to the latter. The opposite fold determines this weight, and fold $k$ evaluates the held-out combination. For sensitivity-ten rating sums, \aes uses the private-pilot configuration in Section~\ref{sec:effectiveness} to compare the same soft blend with Ind-RR before execution.

Algorithm~\ref{alg:aes} summarizes the sensitivity-one branch. The next subsections define the conditional estimator, its L2-loss prediction, and the weight rule.

\begin{algorithm}[t]
\caption{\aes for sensitivity-one statistics}\label{alg:aes}
\KwIn{Public domain $[0,B]$, sensitivity $\Delta=1$, budget $\varepsilon$, range workload $\mathcal Q$}
\KwOut{Estimated counts for all $q\in\mathcal Q$}
\tcp{Vertex side}
Each vertex sends one full-budget geometric report\;
\tcp{Curator side}
Partition reports into two folds using a public hash\;
\ForEach{fold $k$}{
Fit EM on the other fold and rescale to mass $n$\;
\ForEach{$q\in\mathcal Q$}{
Compute predicted coefficients by~\eqref{eq:risk-coefficients}\;
Compute the uncertainty penalty and weight by~\eqref{eq:lambda}\;
Compute $G_{k,q}$ and $L_{k,q}$ from fold $k$\;
}}
\ForEach{$q\in\mathcal Q$}{
Return $\sum_k[(1-\lambda_{k,q})G_{k,q}+\lambda_{k,q}L_{k,q}]$\;
}
\end{algorithm}

\noindent\textbf{Adaptive estimation pipeline.}
\aes transforms one geometric transcript into query-specific estimates through three stages. Cross-fitting first separates histogram fitting from fold execution. The opposite-fold EM fit then supplies the bias, variance, and covariance coefficients used to choose each combination weight. Finally, held-out \bcri and Deg-Lap-RB estimates are blended and summed across folds. The entire adaptation is postprocessing of the original full-budget transcript.

\subsection{Shared-Report Estimation and L2-Loss Prediction}
\bcri uses inverse weights $g_q$ to obtain an unbiased estimate. \aes also uses a second weight vector $\ell_q$ that provides a smoother estimate from the same report. Let $p_L(s)$ be the integer-boundary Laplace hit probability in Section~\ref{sec:degree}, at the same effective noise rate as the geometric report, and define $\ell_q=P^{-1}p_L$. With $r_\alpha=\sqrt\alpha/[2(1+\sqrt\alpha)^2]$ and $f_t=f_q(t)$, the conditional weights are
\begin{equation}\label{eq:lap-rb}
\ell_q(t)=\begin{cases}
f_0+(1-\alpha)r_\alpha[f_1-f_0],&t=0,\\
f_t+r_\alpha[f_{t-1}+f_{t+1}-2f_t],&0<t<B,\\
f_B+(1-\alpha)r_\alpha[f_{B-1}-f_B],&t=B.
\end{cases}
\end{equation}
We call $\widehat C_{LR}=\sum_u\ell_q(T_u)$ \emph{Deg-Lap-RB}. It obtains a conditional Laplace estimate directly from the geometric transcript, so both \aes components share the same private release.

\begin{theorem}\label{thm:lap-rb}
Deg-Lap-RB has the same expectation as integer-boundary Deg-Lap, and no larger variance or expected L2 loss, for each fixed query and collection of statistics.
\end{theorem}
\begin{proof}
Applying the geometric inverse filter to the Laplace density yields an independent jitter $J$ supported on $[-1,1]$. Convolving $J$ with~\eqref{eq:geom} gives Laplace noise with rate $\beta=\varepsilon/\Delta$, and $\Pr[J>1/2]=\Pr[J<-1/2]=r_\alpha$. Taking the conditional expectation of the Laplace hit indicator given the bounded geometric report yields~\eqref{eq:lap-rb}, including the boundary factors induced by clipping. The expectation is preserved, the law of total variance gives no larger variance, and equal bias gives the expected-L2-loss comparison.
\end{proof}
Deg-Lap-RB trades thresholding bias for variance reduction through conditional averaging. It is especially effective when signed thresholding errors are small and the inverse estimator has high variance.

Let $h_s$ be the number of vertices with statistic $s$. For $\lambda\in[0,1]$, define $\widehat C_\lambda=(1-\lambda)\widehat C_G+\lambda\widehat C_{LR}$. With $V_G=\operatorname{Var}(\widehat C_G)$, $V_L=\operatorname{Var}(\widehat C_{LR})$, $C_{GL}=\operatorname{Cov}(\widehat C_G,\widehat C_{LR})$, and $b_L=\mathbb E\widehat C_{LR}-C_q$, the exact expected L2 loss is
\begin{equation}\label{eq:blend-risk}
R_h(\lambda)=(1-\lambda)^2V_G+\lambda^2(V_L+b_L^2)
 +2\lambda(1-\lambda)C_{GL}.
\end{equation}
The covariance is essential because both estimates use the same reports. The coefficients follow from the channel, with products and squares of weight vectors interpreted pointwise:
\begin{align}\label{eq:risk-coefficients}
V_G&=\sum_s h_s[(Pg_q^2)_s-f_q(s)],\nonumber\\
V_L&=\sum_s h_s[(P\ell_q^2)_s-p_L(s)^2],\nonumber\\
C_{GL}&=\sum_s h_s[(P(g_q\ell_q))_s-f_q(s)p_L(s)],\\
b_L&=\sum_s h_s[p_L(s)-f_q(s)].\nonumber
\end{align}
The true histogram is unavailable, so \aes replaces it with a histogram fitted from privatized report counts. It applies standard expectation-maximization (EM), initializing with uniform mass $N$ and normalizing every iterate to $N$. For sensitivity-one statistics, \aes performs twenty iterations on the opposite fold, with $N=n_{-k}$, and then rescales the fitted histogram to mass $n$. Forward and backward recurrences take $O(B)$ time per iteration without forming a dense channel matrix. All fitting is postprocessing.

\subsection{Cross-Fitted Weight Selection}
Let $\widehat V_G,\widehat V_L,\widehat C_{GL},\widehat b_L$ be the predicted coefficients, with their dependence on $q$ suppressed. \aes augments the squared-bias term by $\sigma_k^2=(n/n_{-k})\max(0,\widehat V_G+\widehat V_L-2\widehat C_{GL})$. This model-based penalty accounts for fitting-fold uncertainty in the squared-bias term used for weight selection.

\aes minimizes~\eqref{eq:blend-risk} using the fitted coefficients and replacing $\widehat b_L^2$ by $\widehat b_L^2+\sigma_k^2$. For a nonzero denominator, the combination weight is
\begin{equation}\label{eq:lambda}
\lambda_{k,q}=\left[
\frac{\widehat V_G-\widehat C_{GL}}
{\widehat V_G+\widehat V_L-2\widehat C_{GL}+\widehat b_L^2+\sigma_k^2}
\right]_0^1.
\end{equation}
A zero denominator uses $\lambda_{k,q}=0$. The final estimate sums $(1-\lambda_{k,q})G_{k,q}+\lambda_{k,q}L_{k,q}$ over folds.

For fold error $E_{k,q}=(1-\lambda_{k,q})G_{k,q}+\lambda_{k,q}L_{k,q}-C_{q,k}$, conditioning on the opposite fold fixes the weight, so~\eqref{eq:blend-risk} gives the conditional fold second moment. At aggregate level, the cross moment $2\mathbb E[E_{1,q}E_{2,q}]$ captures the dependence induced by cross-fitted weights.

\subsection{Privacy and Complexity}
\begin{theorem}\label{thm:aes-privacy}
With total budget $\varepsilon$, both \aes configurations satisfy per-vertex $\varepsilon$-edge LDP. For the pilot configuration, a $Q$-query workload satisfies this bound when $\varepsilon_0+Q\varepsilon_1\leq\varepsilon$.
\end{theorem}
\begin{proof}
For sensitivity one, fold assignment, fitting, weight selection, and every query answer are postprocessing of one $\varepsilon$-private geometric report. For the pilot branch, the pilot is $\varepsilon_0$-private. Conditional on any pilot transcript, the chosen execution randomizer is $\varepsilon_1$-private, using either geometric noise calibrated to public sensitivity $\Delta$ or RR on a sensitivity-one bit. Adaptive composition gives $\varepsilon_0+\varepsilon_1$ for one query and at most $\varepsilon_0+Q\varepsilon_1$ for a workload. Fitting, prediction, and selection consume no additional privacy budget.
\end{proof}
With fixed fold count and iteration limit $I$, sensitivity-one \aes uses $O(n+IB+Q\log B)$ arithmetic operations and $O(n+B+Q)$ space in the implementation. Each vertex sends one encoded integer. The pilot branch additionally fits the public checkpoint set. It sends one pilot integer per vertex, reuses one execution geometric integer across blend queries when needed, and adds one RR bit per vertex for each query routed to Ind-RR.

\section{Experimental Evaluation}\label{sec:exp}
% We evaluate accuracy and efficiency on seven degree datasets, followed by a case study of ten public-label degree and signed-rating tasks.

\subsection{Experimental Setup}
\begin{table}[t]\centering\scriptsize
\caption{\add{Summary of Datasets}}\label{tab:summary}
\setlength{\tabcolsep}{3.2pt}
\begin{tabular}{llrrr}
\toprule
\multirow{8}{*}{\textit{Degree datasets}} & Dataset & $n$ & $m$ & $d_{\max}$\\\cmidrule{2-5}
& WikiVote (VT) & 8,298 & 100,762 & 1065\\
& Facebook (FB) & 22,470 & 170,823 & 709\\
& WikiTalk (WT) & 2,394,385 & 4,659,565 & 100029\\
& Petster-Dog (PD) & 451,710 & 8,543,549 & 46503\\
& Google+ (GP) & 107,614 & 12,238,285 & 20127\\
& LiveJournal (LJ) & 4,036,538 & 34,681,189 & 14815\\
& Orkut (OG) & 3,072,627 & 117,185,083 & 33313\\
\midrule
\multirow{5}{*}{\textit{Case study}} & Source (tasks) & $n$ & $m$ & $\Delta$\\\cmidrule{2-5}
& Bitcoin Alpha (2) & 3,783 & 24,186 & 10\\
& Bitcoin OTC (2) & 5,881 & 35,592 & 10\\
& Facebook labels (4) & 22,470 & 170,823 & 1\\
& GitHub labels (2) & 37,700 & 289,003 & 1\\
\bottomrule\end{tabular}

\vspace*{-1em}
\end{table}
\noindent\textbf{Datasets.}
Table~\ref{tab:summary} lists public communication, social, collaboration, and Internet network benchmarks, with up to 4.04 million vertices and 117.19 million edges. In Table~\ref{tab:summary}, $d_{\max}$ is the maximum ordinary degree, whereas $\Delta$ is the public bound on the change in a case-study statistic when one incident edge is added or removed. The ordinary degree graphs are taken from the SNAP~\cite{leskovec2014snap}, except WikiTalk and Petster-Dog, which are taken from KONECT~\cite{kunegis2013konect}. The case study uses public Facebook and GitHub vertex labels~\cite{facebookdata,githubdata} and Bitcoin Alpha/OTC edge ratings~\cite{kumar2016edge}.

\noindent\textbf{Algorithms.}
We compare \bcri and \aes with Deg-RR, Deg-Lap, Ind-Lap, Ind-RR, HH-OUE~\cite{cormode2019range}, HaarHRR~\cite{cormode2019range}, AHEAD~\cite{du2021ahead}, and SquareWave-EMS (SWE)~\cite{li2020squarewave}. HH-OUE uses a four-way hierarchy and consistency reconstruction. HaarHRR uses a padded power-of-two domain. AHEAD uses a binary adaptive tree and an OUE variance-based split threshold. SWE uses 128 bins and forty smoothed EM iterations. The external baselines provide arbitrary-value LDP: their local randomizers accept numerical values throughout the public domain and guarantee $\varepsilon$-LDP for any pair of possible input values. Therefore, if adding or removing one edge changes a vertex's statistic, the two resulting values are protected under the same guarantee.

\noindent\textbf{Parameter settings and evaluation protocol.}
We vary $\varepsilon$ from 0.5 to 3 in steps of 0.5, with default $\varepsilon=2$. Integer query offsets are 0.1\% and 1\% of the benchmark degree span. For ordinary degree, each offset has twenty fixed starts and one hundred stratified random starts. Queries and the ordinary-degree domain are public. \aes uses two folds and twenty EM iterations for sensitivity-one statistics, and default figures use fixed queries with 0.1\% offset. For all experiments, we set the number of testing rounds $T = 10$ and report the average MSE. For each query and method, we average raw squared count errors over independent repetitions and then over queries. Task summaries weight query--budget pairs equally. Isolated timings include report simulation and curator computation but exclude data loading. Communication measures encoded payloads without transport headers. Experiments use C++ on a Linux machine with Intel Xeon Gold 6342 processors (2.80 GHz) and 503 GB RAM.

\subsection{Overall Performance}
\begin{figure}[t]\centering\includegraphics[width=\columnwidth]{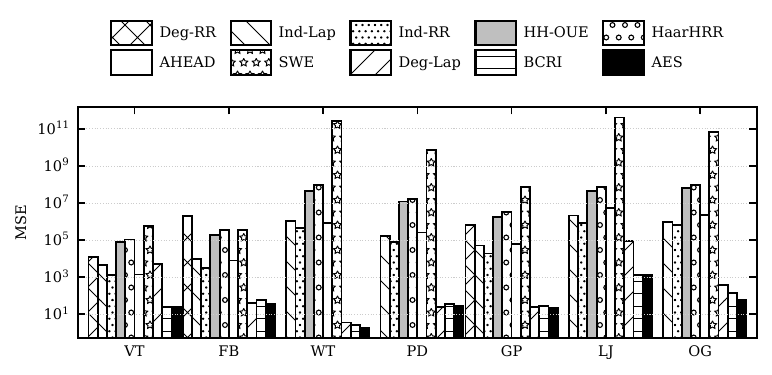}
\vspace*{-1em}
\caption{\add{The mean squared error with $0.1\%\dmax$}}\label{fig:default-all}
\vspace*{-1em}
\end{figure}

\noindent\textbf{Evaluate the effectiveness of all algorithms across different datasets.}
Figure~\ref{fig:default-all} compares the algorithm families at $\varepsilon=2$ with fixed queries and 0.1\% offset. The Facebook results demonstrate the complementary benefits of direct inversion and adaptive estimation. \bcri attains MSE 57.05, while Deg-Lap attains 40.26. \aes reduces MSE to 36.03, which is 36.85\% below \bcri and 10.49\% below Deg-Lap. Its private fit identifies a favorable balance between \bcri's unbiased inverse and the lower-variance conditional estimator. Other algorithm families show distinct bottlenecks. Ind-RR attains 3,222.19 because every vertex contributes population-wide indicator variance, while AHEAD attains 8,115.05 as it reconstructs a hierarchy over the domain rather than directly targeting the query endpoints. These comparisons show how numerical locality and query-specific loss prediction address complementary error sources.

\begin{figure}[t]\centering\includegraphics[width=\columnwidth]{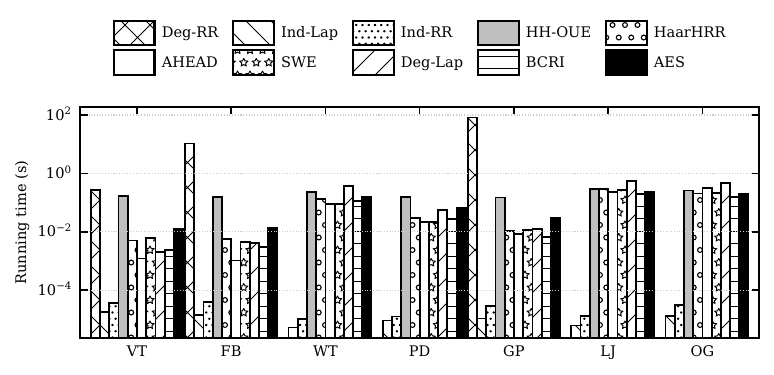}
\vspace*{-1em}
\caption{\add{The computational time cost with $0.1\%\dmax$.}}\label{fig:runtime}
\vspace*{-1em}
\end{figure}

\noindent\textbf{Evaluate the computational time costs across datasets.}
Figure~\ref{fig:runtime} focuses on computational scalability for the default workload. Deg-RR has the highest measured running times and exceeds the practical resource budget on the large shared-graph settings. Its construction randomizes $\Theta(n^2)$ adjacency pairs, making both report generation and degree accumulation prohibitive as the graph grows. The other methods avoid this dense-pair enumeration. Deg-Lap and SWE process scalar reports, indicator methods evaluate only the requested predicates, and hierarchical or wavelet methods use compact domain structures. \bcri and \aes replace randomized-graph construction with one integer report per vertex followed by histogram and prefix computations. \aes therefore completes each reported degree workload with mean running time at most 0.23 seconds, including the largest graphs in the study.

\subsection{Accuracy under Different Query Settings}
\begin{figure}[t]\centering\includegraphics[width=\columnwidth]{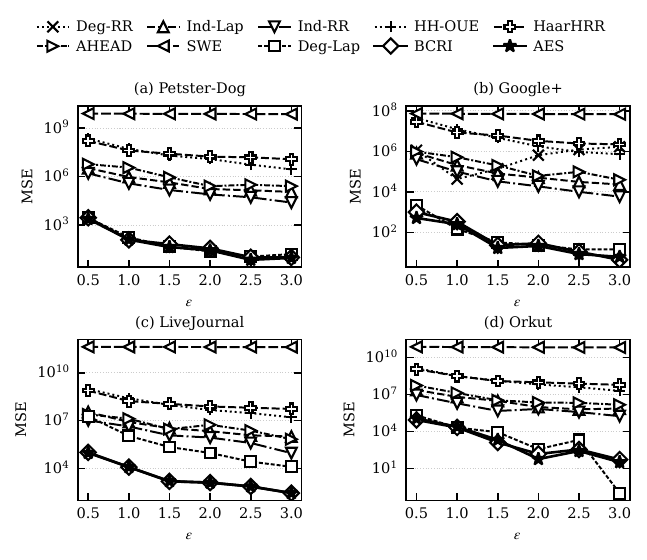}
\vspace*{-2em}
\caption{Fixed degree ranges, 0.1\% offset.}\label{fig:fixed-narrow}
\vspace*{-1em}
\end{figure}

\begin{figure}[t]\centering\includegraphics[width=\columnwidth]{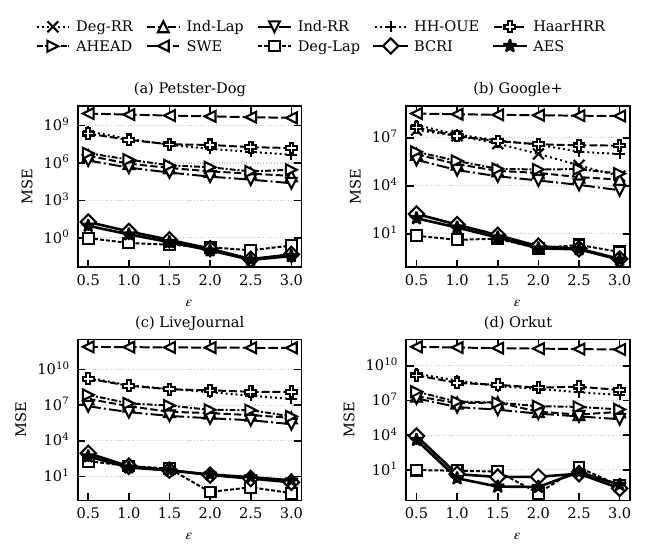}
\vspace*{-2em}
\caption{Fixed degree ranges, 1\% offset.}\label{fig:fixed-wide}
\vspace*{-1em}
\end{figure}

\noindent\textbf{Evaluate the effect of the privacy budget $\varepsilon$.}
Figures~\ref{fig:fixed-narrow} and~\ref{fig:fixed-wide} reveal how the algorithms convert additional privacy budget into accuracy. Ind-Lap and Ind-RR benefit directly through their analytic losses $2n/\varepsilon^2$ and $n e^\varepsilon/(e^\varepsilon-1)^2$. Deg-Lap also receives less scalar noise, although its squared thresholding-bias term can dominate the remaining error. For \bcri, increasing $\varepsilon$ decreases $\alpha=e^{-\varepsilon/\Delta}$ and suppresses inverse amplification. \aes additionally adjusts the combination using the fitted bias, variance, and covariance. On LiveJournal with fixed 0.1\%-offset queries, \aes decreases monotonically from MSE 94,099.35 at $\varepsilon=0.5$ to 283.96 at $\varepsilon=3$, a 99.70\% reduction. This pronounced response demonstrates that \aes effectively exploits the improved report quality at larger privacy budgets.

\begin{figure}[t]\centering\includegraphics[width=\columnwidth]{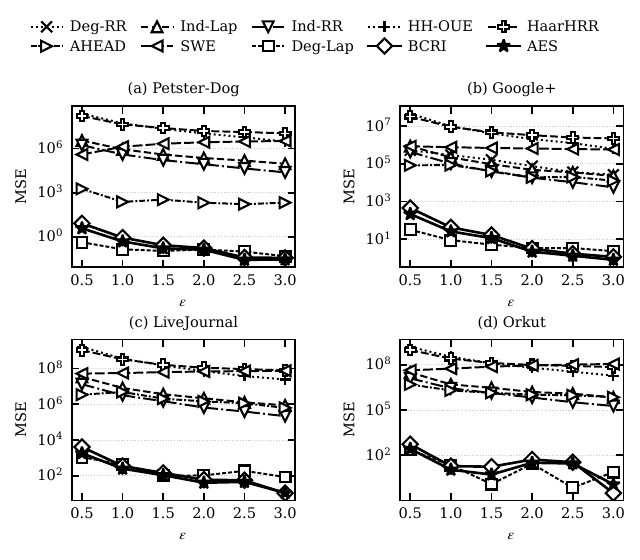}
\vspace*{-2em}
\caption{Random-start degree ranges, 0.1\% offset.}\label{fig:random-narrow}
\vspace*{-1em}
\end{figure}

\begin{figure}[t]\centering\includegraphics[width=\columnwidth]{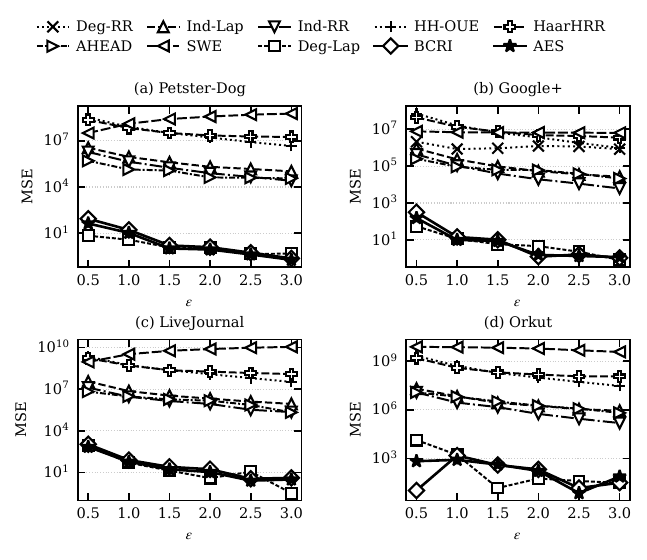}
\vspace*{-2em}
\caption{Random-start degree ranges, 1\% offset.}\label{fig:random-wide}
\vspace*{-1em}
\end{figure}

\noindent\textbf{Evaluate the effect of the query range.}
The offsets and query starts in Figures~\ref{fig:fixed-narrow}--\ref{fig:random-wide} change the statistic mass surrounding the endpoints, creating complementary operating regimes. Deg-Lap and Deg-Lap-RB benefit from rare false positives and low hit-probability variance in sparse tails, while populated endpoints can accumulate signed thresholding errors into substantial squared bias. \bcri eliminates this thresholding bias and concentrates its variance near the endpoints through local inverse weights. \aes predicts both effects and their covariance, then assigns a continuous combination weight to each query. On Orkut with fixed 1\%-offset ranges at $\varepsilon=1.5$, \aes attains MSE 0.37, compared with 2.48 for \bcri and 7.30 for Deg-Lap, corresponding to reductions of 85.18\% and 94.97\%. The example highlights \aes's ability to combine the variance advantage of conditional estimation with the direct range-correction advantage of \bcri as query locations vary.

\subsection{A Fixed Total Workload Budget}
\begin{figure}[t]\centering\includegraphics[width=\columnwidth]{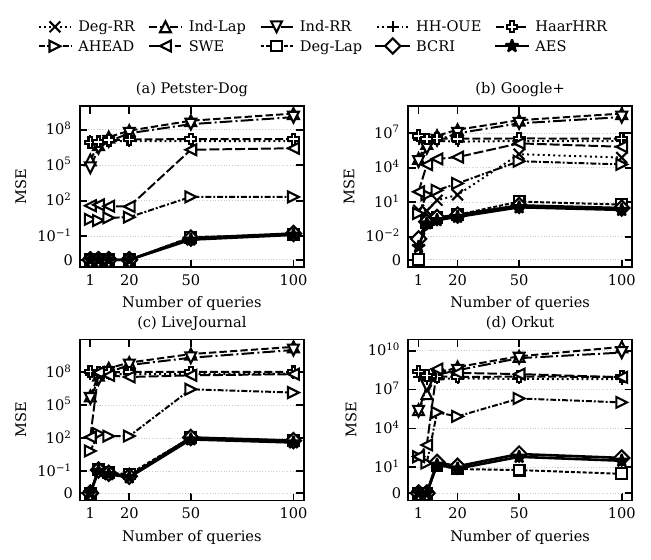}
\vspace*{-2em}
\caption{\add{Random workloads under total budget $\varepsilon=2$ with 0.1\% offset. }}\label{fig:workload}
\vspace*{-1em}
\end{figure}

\noindent\textbf{Evaluate the effectiveness of report reuse under a fixed total workload budget.}
Figure~\ref{fig:workload} fixes the total budget at $\varepsilon=2$ and increases the number of random-start, 0.1\%-offset queries. This experiment isolates workload scaling rather than per-query accuracy alone. Ind-Lap and Ind-RR must allocate $\varepsilon/Q$ to each predicate, so their indicator noise grows rapidly with $Q$. Scalar and full-distribution methods instead reuse one release and retain their reporting noise scale across the workload. \bcri combines this reuse with direct range inversion, while \aes adds query-specific combination. At $Q=100$ on Orkut, \aes attains MSE \limitedreps{31.98}, compared with \limitedreps{52.54} for \bcri and \limitedreps{$7.29\times10^9$} for Ind-RR. \aes reduces MSE by \limitedreps{39.14\%} relative to \bcri, while the contrast with Ind-RR demonstrates the decisive benefit of reusing one private transcript under a fixed workload budget.

\subsection{Scalability and Communication}
\begin{figure}[t]\centering\includegraphics[width=\columnwidth]{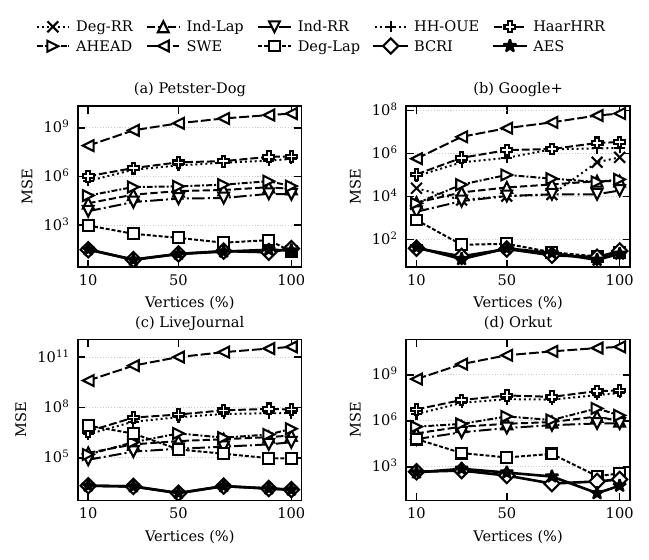}
\vspace*{-1em}
\caption{Induced-subgraph scalability in the default setting.}\label{fig:scaling}
\vspace*{-1em}
\end{figure}

\noindent\textbf{Evaluate the effect of the number of vertices.}
Figure~\ref{fig:scaling} uses induced subgraphs with 10\%, 30\%, 50\%, 70\%, and 90\% of the vertices, followed by the full graph, and recomputes degrees after sampling. The indicator methods have population-wide variance terms proportional to $n$, and full-distribution methods reconstruct estimates across their domain structures. Deg-RR additionally incurs dense-pair report generation. \bcri and \aes use a single reporting pass followed by histogram-based query evaluation, and maintain lower MSE than the indicator and distribution-estimation methods from the sampled subgraphs to the full graphs. Their accuracy and computational structure therefore scale together with the vertex population.

\begin{figure}[t]\centering\includegraphics[width=\columnwidth]{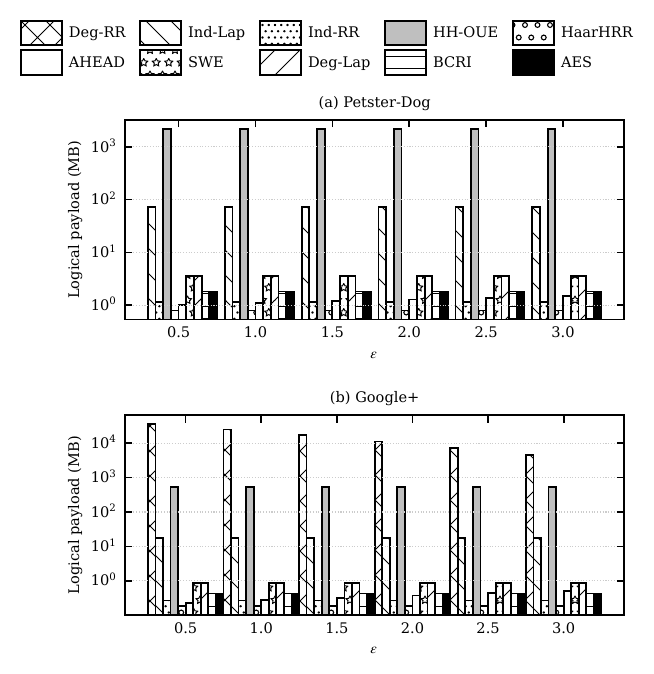}
\vspace*{-1em}
\caption{Logical communication for twenty fixed queries with 0.1\% offset.}\label{fig:communication}
\vspace*{-1em}
\end{figure}

\noindent\textbf{Evaluate the communication costs of all algorithms under $\varepsilon$-edge LDP.}
Figure~\ref{fig:communication} compares logical upload sizes for an entire workload of twenty fixed, 0.1\%-offset queries. \bcri and \aes require one 32-bit integer per vertex, independent of the number of queries. Deg-Lap and SWE use 64-bit scalar reports, indicator methods transmit new predicate reports as the workload grows, and hierarchical methods encode level information. On Google+ at $\varepsilon=2$, \aes uploads 0.43 MB, compared with 17.22 MB for Ind-Lap and 521.34 MB for HH-OUE. Report reuse thus provides a substantial communication advantage in addition to compact per-vertex encoding.

\subsection{Component Analysis}
\noindent\textbf{Evaluate the effect of boundary conditioning in \bcri.}
\add{Table~\ref{tab:boundary-ablation} compares the bounded and unbounded inverses over all query groups and budgets. Conditioning reduces MSE by 92.34\% on Petster-Dog, 79.82\% on GitHub label 1, and 55.84\% on Google+. Replacing unbounded tail contributions by their conditional expectations removes uninformative randomness. Theorem~\ref{thm:geo-rb} proves the loss guarantee. This provides an explicit clipped-report conditional estimator beyond Gu et al.~\cite{gu2019supporting}.}
\begin{table}[t]\centering\footnotesize
\caption{\add{Boundary conditioning. Positive gain favors \bcri.}}\label{tab:boundary-ablation}
\begin{tabular}{lrrr}
\toprule
Dataset & Unbounded & \bcri & Gain (\%)\\\midrule
WikiVote & 163.29 & 76.49 & 53.15\\
GitHub-L1 & 1,880.31 & 379.49 & 79.82\\
Petster-Dog & 668.72 & 51.23 & 92.34\\
Google+ & 187.22 & 82.67 & 55.84\\
\bottomrule\end{tabular}

\vspace*{-1em}
\end{table}

\noindent\textbf{Evaluate the query-dependent value of \aes.}
\add{For nonempty interior queries with $C_q>0$ and $x>1$, we first average MSE within each task over all query groups and budgets and then equally weight the 15 tasks containing this regime. \aes attains MSE 799.32, compared with 328,091.36 for \bcri and 9,208.92 for Deg-Lap, reducing error by 99.76\% and 91.32\%, respectively. This regime directly demonstrates the value of balancing endpoint variance against thresholding bias through private prediction.}

\noindent\textbf{Evaluate the \aes estimator-combination modules.}
\add{Table~\ref{tab:aes-modules} uses identical reports within every repetition. On Google+ at $\varepsilon=0.5$, full \aes reduces MSE by 48.57\% relative to both \bcri-only and hard selection, and by 63.86\% relative to Deg-Lap-RB-only. Setting $C_{GL}=0$ leaves higher MSE on Facebook, Google+, and WikiTalk, with full-\aes reductions of 21.79\%, 13.75\%, and 15.91\%. Across the six diagnostic settings, predicted loss and realized squared error have Spearman correlations of 0.79--0.92.}
\begin{table}[t]\centering\footnotesize\setlength{\tabcolsep}{3pt}
\caption{\add{\aes module ablation over ten paired repetitions. Gain is the MSE reduction to full \aes.}}\label{tab:aes-modules}
\begin{tabular}{lrrrr}
\toprule
Setting & Ablation & MSE & \add{MSE(\aes)} & Gain (\%)\\\midrule
\multicolumn{5}{l}{\emph{Continuous blending}}\\
GP, $\varepsilon=0.5$ & \bcri-only & 1,009.09 & 519.00 & 48.57\\
GP, $\varepsilon=0.5$ & Deg-Lap-RB-only & 1,435.98 & 519.00 & 63.86\\
GP, $\varepsilon=0.5$ & Hard selection & 1,009.09 & 519.00 & 48.57\\
\midrule
\multicolumn{5}{l}{\emph{Covariance modeling}}\\
FB, $\varepsilon=2$ & \add{w/o $C_{GL}$} & 46.07 & 36.03 & 21.79\\
GP, $\varepsilon=2$ & \add{w/o $C_{GL}$} & 25.14 & 21.69 & 13.75\\
WT, $\varepsilon=1.5$ & \add{w/o $C_{GL}$} & 2.64 & 2.22 & 15.91\\
\bottomrule\end{tabular}

\vspace*{-1em}
\end{table}

\subsection{\add{Paired Comparison and Variance Calibration}}\label{sec:paired}
\add{We evaluate seven graphs under four query groups and six privacy budgets, yielding 168 settings with ten repetitions each. On identical reports, the geometric mean of the per-setting \aes-to-\bcri MSE ratios is 0.78, corresponding to a 22\% reduction. For each graph, we divide the summed $\widehat V_q$ by the summed exact variance over its queries and repetitions. The seven ratios lie between 0.995 and 1.000. We also measure the empirical coverage of the resulting nominal 95\% error bars. Across the resulting 100,800 query records, 98.1\% contain the true count. This coverage is an empirical calibration result, not a finite-sample guarantee.}
\begin{table}[t]\centering\scriptsize\setlength{\tabcolsep}{2.5pt}
\caption{\add{ GM and Max summarize MSE ratios to \bcri; Ratio and Cov. summarize variance estimation.}}\label{tab:paired}
\add{\begin{tabular}{lrrrrrrr}
\toprule
 & \multicolumn{2}{c}{\aes} & RB & \multicolumn{2}{c}{Geo-EM} & \multicolumn{2}{c}{$\widehat V_q$}\\\cmidrule(lr){2-3}\cmidrule(lr){4-4}\cmidrule(lr){5-6}\cmidrule(lr){7-8}
Data & GM & Max & GM & GM & Max & Ratio & Cov.\\\midrule
VT & 0.80 & 1.39 & 2.23 & 0.84 & 1.00 & 0.997 & 96.8\\
FB & 0.95 & 1.38 & 7.16 & 0.91 & 1.03 & 0.999 & 96.1\\
WT & 0.67 & 1.04 & 0.33 & 0.70 & 1.00 & 0.996 & 99.6\\
PD & 0.72 & 1.17 & 0.41 & 0.76 & 1.00 & 0.995 & 99.0\\
GP & 0.72 & 1.13 & 0.51 & 0.87 & 1.00 & 1.000 & 97.7\\
LJ & 0.79 & 1.17 & 1.94 & 0.90 & 1.00 & 1.000 & 98.9\\
OG & 0.86 & 1.98 & 0.99 & 0.90 & 1.00 & 1.000 & 98.4\\
\bottomrule\end{tabular}
}
\end{table}

\subsection{Case Study: Other Vertex Statistics}\label{sec:effectiveness}
\noindent\textbf{Case-study tasks and parameter settings.}
\add{The case study contains six public-label-degree tasks on Facebook and GitHub and four incoming/outgoing signed-rating tasks on Bitcoin Alpha and OTC. Public-label degrees have sensitivity one and domains $[0,4097]$ and $[0,65536]$, respectively. Ratings have sensitivity ten, are shifted by 4096, and use domain $[0,8193]$. Each task uses twenty fixed and twenty random starts per offset. Public-label tasks use full-budget cross-fitting. Rating tasks use a 15\% pilot and 85\% execution budget, evaluate EM checkpoints $\{20,80,200,1000,5000\}$, minimize the maximum predicted blend loss, and select Ind-RR on ties.}

\begin{figure}[t]\centering\includegraphics[width=\columnwidth]{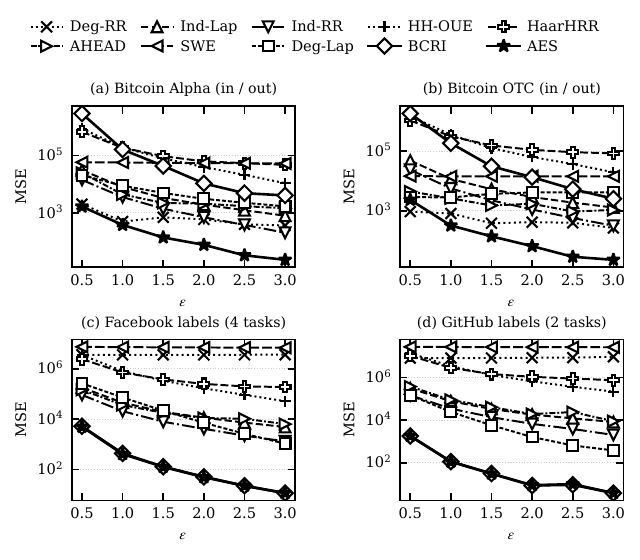}
\vspace*{-1em}
\caption{\add{MSE on public-label-degree and signed-rating tasks.}}\label{fig:effectiveness}
\vspace*{-1em}
\end{figure}

\noindent\textbf{Evaluate the effectiveness of \bcri and \aes on other vertex statistics.}
\add{Figure~\ref{fig:effectiveness} evaluates the same interface across ordinary degree, public-label degree, and signed-rating statistics. Deg-RR shares randomized graph releases by source, using undirected adjacency for public labels and directed adjacency for ratings. On Bitcoin Alpha incoming ratings, \bcri, Deg-Lap, and Ind-RR attain MSEs 359,311.57, 10,554.97, and 3,345.84. Sensitivity ten amplifies numerical perturbation, whereas Ind-RR protects a sensitivity-one indicator. \aes uses predicted loss to select the execution and attains MSE 217.68, reducing error by 99.94\% from \bcri and 93.49\% from Ind-RR. Its reduction over \bcri exceeds 99.8\% on all four rating tasks.}

\noindent\textbf{Evaluate the effect of the \aes pilot-budget allocation.}
\add{Table~\ref{tab:pilot-budget} varies the pilot fraction $\rho$ on the four rating tasks. Mean MSE is 347.20 at 5\%, 147.27 at 10\%, 143.78 at 15\%, and 196.50 at 20\%. A small pilot weakens histogram fitting, while an excessive pilot reduces the execution budget. The 15\% split achieves the lowest mean and is used by default.}
\begin{table}[t]\centering\footnotesize
\caption{Pilot-budget fractions on rating sums. The execution receives the remaining budget.}\label{tab:pilot-budget}
\begin{tabular}{lrrrr}
\toprule
$\varepsilon$ & 5\% & 10\% & 15\% & 20\%\\\midrule
1 & 960.13 & 361.61 & 340.53 & 481.57\\
2 & 64.05 & 59.46 & 66.74 & 82.72\\
3 & 17.42 & 20.75 & 24.06 & 25.22\\
Mean & 347.20 & 147.27 & 143.78 & 196.50\\
\bottomrule\end{tabular}

\vspace*{-1em}
\end{table}

\section{Related Work}
\noindent\textbf{\add{Graph analytics and range computation.}}
\add{Scalable graph systems address biclique counting, enumeration, and temporal community computation through task-specific structural optimizations~\cite{yang2021biclique,wang2023uncertainBiclique,chen2022maximalBiclique,chen2026biclique}. Parallel, streaming, and distributed techniques support shortest-path counting, triangle counting, and subgraph matching on large graphs~\cite{wang2023shortestPathCounting,gou2021sliding,pandey2021trust,wang2019rdfSubgraph}.}

\noindent\textbf{\add{Private graph statistics and range queries.}}
\add{Central differential privacy assumes a trusted curator~\cite{dwork2014algorithmic,jiang2021applications}, whereas edge LDP randomizes local information before transmission~\cite{imola2021locally,imola2022communication}. Local graph methods reconstruct graphs or summaries for degree and subgraph estimation~\cite{qin2017generating,zhang2018two,gao2018local,ye2020lf}, while node-DP projection and smooth sensitivity provide complementary central-model controls~\cite{kasiviswanathan2013analyzing,day2016publishing,macwan2018node,nissim2007smooth}. Hierarchical, Haar, and numerical-distribution summaries support reusable local range estimates~\cite{cormode2019range,wang2017oue,du2021ahead,li2020squarewave}. Privacy-preserving graph analytics also cover anonymization, outsourced computation, link prediction, and federated learning~\cite{wu2010ksymmetry,gao2013outsourcing,ran2024privateGNN,xu2025sectric,wang2020federatedLDA}. Gu et al.~\cite{gu2019supporting} use distance-based reports and probability-matrix inversion for unbiased frequency and range estimation, and geometric mechanisms support query-specific postprocessing~\cite{ghosh2012geometric}. \bcri develops bounded geometric range channels with local inversion, constant-time queries, boundary conditioning, and endpoint variance estimation. \aes privately predicts bias, variance, and covariance for shared-report combination.}

\section{Conclusion}
We studied vertex-statistic range counting under edge local differential privacy and developed a layered framework spanning scalar statistics, range indicators, bounded-channel inversion, and adaptive estimation. Deg-RR and Deg-Lap characterize complementary statistic-estimation costs, while Ind-Lap and Ind-RR provide direct query-specific estimates. \bcri derives an explicit local inverse and a boundary-conditioned estimator with a proved expected-L2-loss guarantee, supporting constant-time queries from one reusable transcript. \aes forms the adaptive core by predicting bias, variance, and covariance to combine estimators for each query. \add{Experiments on diverse graph tasks demonstrate accuracy, efficiency, compact communication, and workload scalability.}

\section*{AI-Generated Content Acknowledgment}
OpenAI GPT assisted with code inspection, experiment orchestration, mathematical analysis, figure preparation, and language polishing. The authors reviewed and verified all algorithm designs, theoretical results, experiments, and manuscript content, and take full responsibility for the correctness, integrity, and originality of the paper.\par

\bibliographystyle{IEEEtran}
\bibliography{sample}
\end{document}